# Vibration Combined High Temperature Cycle Tests for Capacitive MEMS Accelerometers

Z. Szűcs, G. Nagy, S. Hodossy, M. Rencz, A. Poppe
Budapest University of Technology and Economics, Department of Electron Devices
H-1521 Budapest, Goldmann György tér 3., Hungary
E-mail: <szucs|nagyg|hodossy|rencz|poppe>@eet.bme.hu

*Abstract -* In this paper vibration combined high temperature cycle tests for packaged capacitive SOI-MEMS accelerometers are presented. The aim of these tests is to provide useful Design for Reliability information for MEMS designers. A high temperature test chamber and a chopper-stabilized read-out circuitry were designed and realized at BME – DED. Twenty thermal cycles of combined Temperature Cycle Test and Fatigue Vibration Test has been carried out on 5 samples. Statistical evaluation of the test results showed that degradation has started in 3 out of the 5 samples.

## I. INTRODUCTION

Micromechanical sensors are more and more widespread in today's electrical sensing applications. A considerable part of these sensors uses capacitive sensing elements. The big advantages of these are the low current consumption and high resolution. Another particular advantage of capacitive sensors is the ease of integration with the surrounding circuitry [1].

Since MEMS devices contain mechanical elements, the reliability and lifetime of these structures is a big concern. The aim of reliability testing is to provide useful Design for Reliability information for MEMS designers.

## II. TEMPERATURE CYCLE TESTS

Temperature Cycle Testing (TCT) or simply temperature cycling, determines the ability of parts to resist extremely low and extremely high temperatures, as well as their ability to withstand cyclical exposures to these temperatures. A mechanical failure resulting from cyclic thermomechanical load is known as a fatigue failure, so temperature cycling primarily accelerates fatigue failures.

Failure acceleration due to Temperature Cycling and Thermal Shock depends on the following factors:
- the difference between the high and low temperatures used
- the transfer time between the two temperatures
- the dwell times at the extreme temperatures.

Failure mechanisms accelerated by temperature cycling include die cracking, package cracking, neck/heel/wire breaks, and bond lifting.

For reliability testing or qualification of new devices, 1000 temp cycles are usually performed, with interim visual inspection and electrical test read points at 200X and 500X.

Two industry standards that govern Temperature Cycle Testing are the MIL-STD-883 Method 1010 [5] and the JEDEC JESD22-A104.

Higher failure acceleration can be achieved when thermal and mechanical effects are present simultaneously. For this reason we combine vibration fatigue tests with temperature cycle tests.

## III. THE DEVICE UNDER TEST

The device under test is a SOI-MEMS accelerometer produced and provided by Qinetiq. This comb-drive structure is built up of one movable and two static plates creating two differential capacitors, a capacitive half-bridge. The displacement of the moving electrode results in a change in the capacitance.

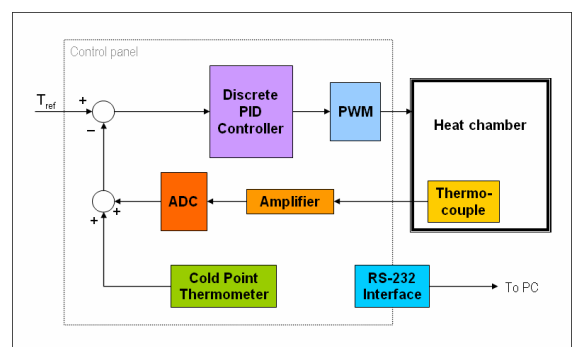

Fig. 1. Block diagram of thermal test chamber

## IV. TEST EQUIPMENT

A high temperature test chamber was developed at BME – Department of Electron Devices. The device is able to produce temperature values in the range of 0…250 °C (±1°C) with the help of a digital PID controller. The temperature is sensed by a thermocouple and the heating of the chamber is PWM modulated. Fig. 1 shows the block diagram of the system.

Electrical testing of the samples to device specifications shall also be performed to detect electrical failures accelerated by the temperature cycles. For this reason a chopper-stabilized read-out circuitry has been built of discrete elements at BUTE (Fig. 2).

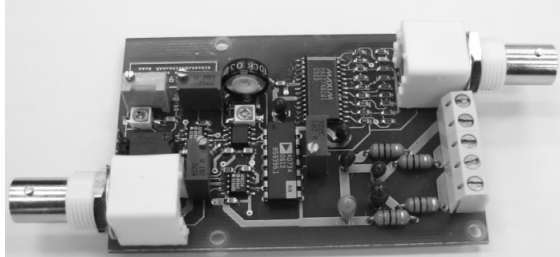

Fig. 2. Chopper-stabilized readout circuitry for electrical characteriziation of capacitive MEMS sensors

## V. Test parameters

Our test consists of the following four main phases:
⇒ Initial electrical characterization of the samples at ±1g static acceleration before testing.
⇒ Combined TCT-FVT test according to the given test specification.
⇒ Electrical characterization of the samples at ±1g static acceleration after 20, 50 and 100 thermal cycles.
⇒ Statistical analysis of the data (Welch's t-test)

### A. Initial characterization

Several initial characterising measurements have been done on five different SOI-MEMS accelerometers at BUTE. Our primary intention was to gain a quantitative characterization of the devices before the accelerated lifetime test to allow for later comparison. We accomplished a series of measurements on the devices at a constant excitation and observed the results of the different devices within each sequence and the results of each device across the series. The former is the indicator of the stability of the sensor process technology, the latter gives information on the stability of the sensors, i. e. their quality of producing an invariant response to a constant excitation.

Figure 3 and 4 show the summarized results of the samples. The average output values were 60.843 mV for +1g and -52.475 mV for -1g, the standard deviation were 1.249 mV (1.99 %) and 1.68 mV (3.23 %) respectively.

The small values of the deviation show that the measurement and the devices were stable.

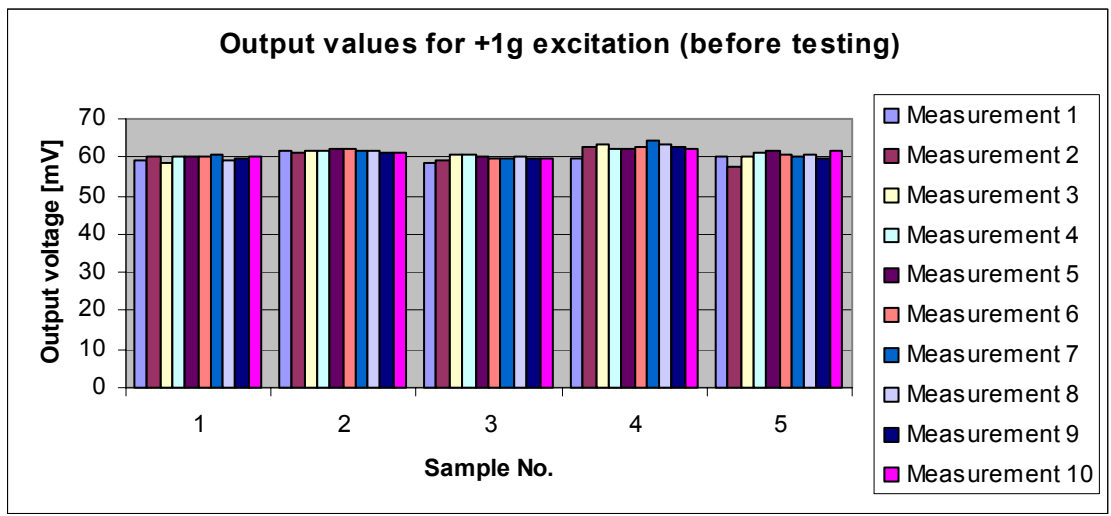

Fig. 3. Output values for +1g excitation (before testing) [18 May 2007]

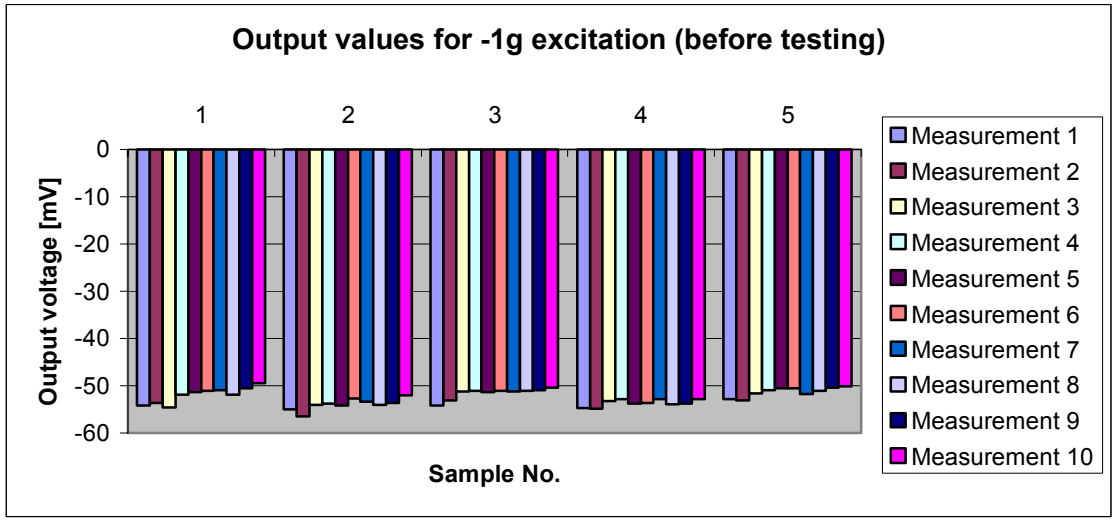

Fig. 4. Output values for -1g excitation (before testing) [18 May 2007]

## VI. COMBINED TCT-FVT TEST

### A. TCT parameters

For the combined TCT-FVT test the desired temperature waveform is shown in Fig. 5. One thermal cycle consists of an initial storage period (called dwell time) at a temperature of TL, a warm-up period with a specified slope from TL to TH, a high temp storage period at TH, and finally a cool-down period from TH back to TL with the same slope as specified for warming up.

In our test the temperature is cyclically changing between TL = 50 °C and TH = 150 °C. The dwell times for both low and high temperature states (TL and TH) are 5 minutes while the rise and fall times are 7 minutes (this results in a temp slope of $\partial T/\partial t$ = 0.24 °C/sec = 14.4 °C/min).

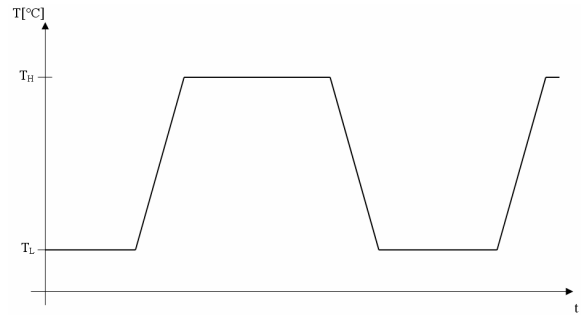

Fig. 5. Temperature waveform for combined TCT-FVT testing

Fig. 6 shows two periods of the measured temperature waveform inside the heat chamber during testing. As it can be seen, the temperature followed the desired waveform of Fig. 5 very accurately.

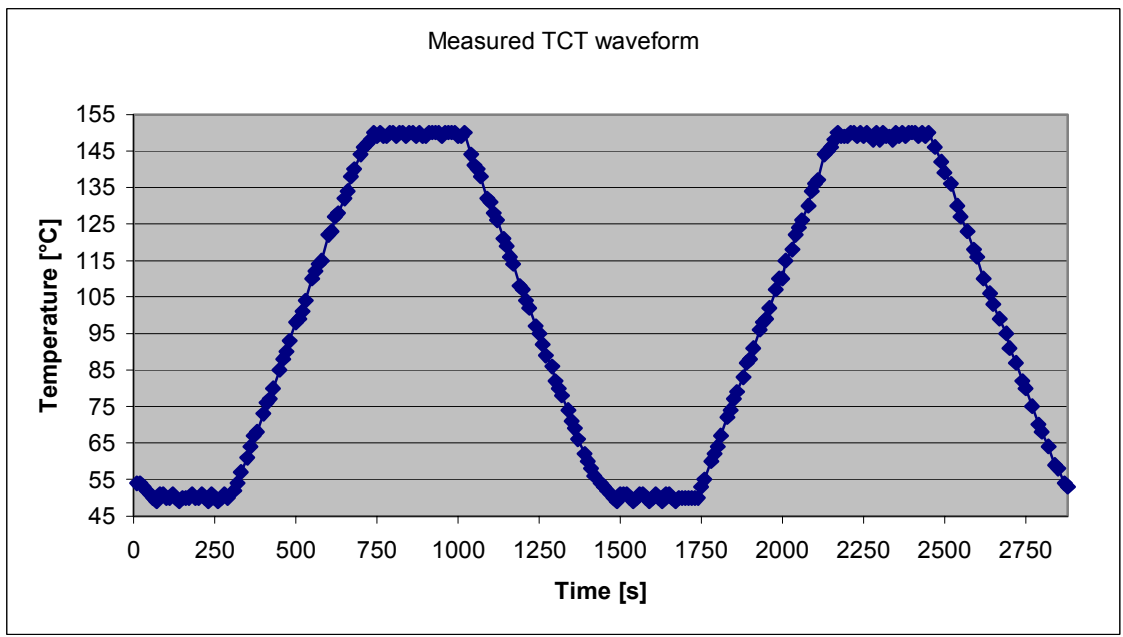

Fig 6. The measured TCT waveform during testing.

## B. FVT parameters

As the devices under test are capacitive sensors, we ran MATLAB simulations to see if there was a possibility to vibrate the movable electrode electronically. The input parameters for the FVT test simulation were the following:

⇒ The peak acceleration of the movable electrode should be in the vicinity of $|a_{max}| = 70g$.

⇒ The highest deflection should be less than the sense gap: 5 μm.

⇒ The frequency of the excitation signal ($f_e$) should be near the half of the resonance frequency value ($f_{res}$ = 2.37 kHz) of the device (as the interaction between the electrodes is always attractive, the frequency of the mechanical vibration will be $2f_e=f_{res}$).

⇒ The excitation waveform should be a harmonic sinewave.

Our simulation results showed that an excitation signal of $U_p$=14.5 V amplitude and $f_e$=950 Hz frequency would result in a peak acceleration of $a_{max}$= 73.733g and a total deflection of 4.844 μm. We applied this excitation as an FVT signal to the samples being under TCT test in the heat chamber.

## VII. ELECTRICAL CHARACTERIZATION AFTER 20 TCT CYCLES

20 periods of combined TCT-FVT test were applied to the samples and the same characterising measurements have been performed on the devices as described in Section V.

Fig. 7 and 8 show the summarized results of the different samples. The average output values were 62.598 mV for +1g and -51.412 mV for -1g, the standard deviation was 1.712 mV (2.7 %) and 1.867 mV (3.66 %) respectively.

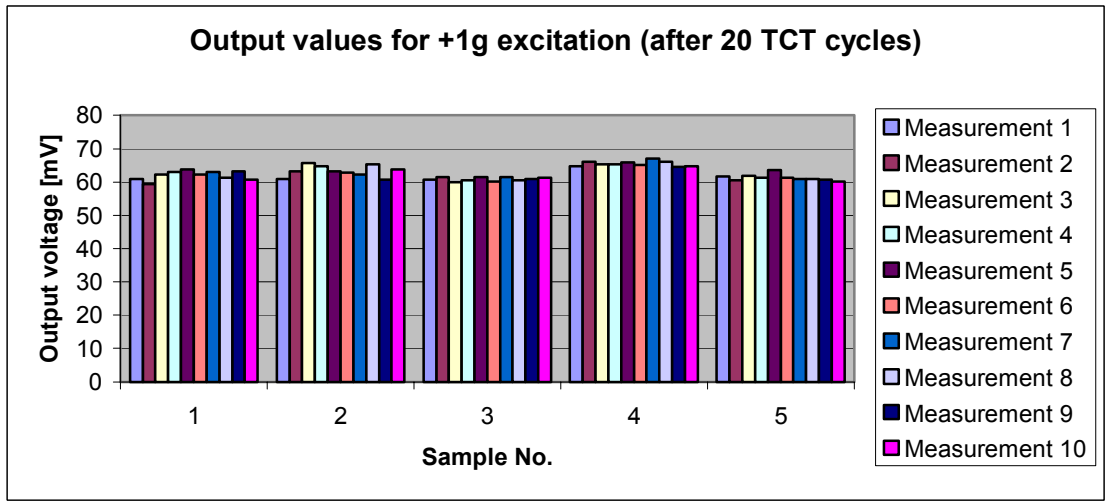

Fig. 7. Output values for +1g excitation (after 20 TCT cycles) [18 May 2007]

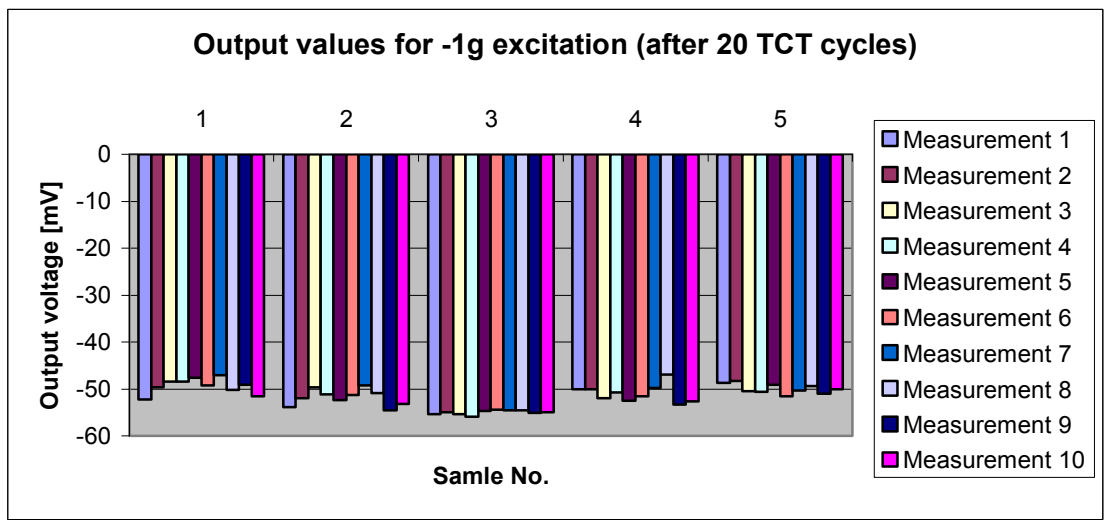

Fig. 8. Output values for -1g excitation (after 20 TCT cycles) [18 May 2007]

## VIII. Statistical analysis of the data (Welch's t-test)

As the difference between the average output values before and after the 20 cycle test is very small, it is not evident if there was any degradation in the samples. In such cases the Welch's t-test is a powerful statistical method to analize the results. If the calculated *t* value is above the threshold chosen for statistical significance (usually the 0.05 level), then the null hypothesis that the two groups do not differ is rejected in favor of an alternative hypothesis stating that the two groups *do* differ. Table 1 summarizes the results of the test based on our measured data.

| No. | Output values for -1g | | | | Welch's t-test | | | | |
|---|---|---|---|---|---|---|---|---|---|
| | Before test | | After 20 cycles | | $t$ statistic value | $p$ statistical significance | $f$ No. of degrees of freedom | $tp$ value | Result |
| | Mean [mV] | Std. dev. [mV] | Mean [mV] | Std. dev. [mV] | | | | | |
| 1 | -51,963 | 1,6800 | -49,375 | 1,6152 | 2,4831 | 0,05 | 7,9876 | 2,37 | FAILED |
| 2 | -53,912 | 1,2116 | -51,833 | 1,7196 | 2,2099 | 0,05 | 7,1861 | 2,37 | OK |
| 3 | -51,571 | 1,1588 | -54,947 | 0,4621 | 6,0512 | 0,05 | 5,2407 | 2,57 | FAILED |
| 4 | -53,622 | 0,7492 | -50,947 | 1,8670 | 2,9734 | 0,05 | 5,2556 | 2,57 | FAILED |
| 5 | -51,307 | 1,0136 | -49,959 | 1,0503 | 2,0650 | 0,05 | 7,9899 | 2,37 | OK |

| No. | Output values for +1g | | | | Welch's t-test | | | | |
|---|---|---|---|---|---|---|---|---|---|
| | Before test | | After 20 cycles | | $t$ statistic value | $p$ statistical significance | $f$ No. of degrees of freedom | $tp$ value | Result |
| | Mean [mV] | Std. dev. [mV] | Mean [mV] | Std. dev. [mV] | | | | | |
| 1 | 59,817 | 0,6643 | 61,998 | 1,3823 | 3,1798 | 0,05 | 5,7542 | 2,57 | FAILED |
| 2 | 61,729 | 0,3649 | 63,283 | 1,7129 | 1,9841 | 0,05 | 4,3624 | 2,78 | OK |
| 3 | 59,815 | 0,6279 | 60,893 | 0,5712 | 2,8398 | 0,05 | 7,9295 | 2,37 | FAILED |
| 4 | 62,495 | 1,2496 | 65,513 | 0,8010 | 4,5466 | 0,05 | 6,8124 | 2,45 | FAILED |
| 5 | 60,36 | 1,1844 | 61,304 | 0,9385 | 1,3969 | 0,05 | 7,6027 | 2,37 | OK |

Table 1. Results of the Welch's t-test

The statistical results for +1g and for -1g were the same. The Welch's t-test showed that some kind of degradation has started in samples: No. 1, 3 and 4.

## IX. Conclusion

In this document the standard TCT and FVT tests, based on the MIL-STD-883F, were presented. With the help of these standard test specifications BUTE determined a combined TCT-FVT test plan for the MEMS devices provided by Qinetiq.

A thermal test chamber and a chopper-stabilized read-out circuitry were also designed, built and characterized in order to carry out the tests.

After the initial characterization steps, 20 thermal cycles of the test has been carried out. Albeit none of the devices suffered total failure, the statistical analysis of the measured data (Welch's t-test) showed that some kind of degradation has started in samples No. 1, 3 and 4. Further cycles of the combined test, presented above, will hopefully result in detectable failure to some of the three structures in question.

## Acknowledgment

This work was supported by the PATENT IST-2002-507255 Project of the EU and by the OTKA-TS049893 project of the Hungarian Government.